# Using Multiple Seasonal Holt-Winters Exponential Smoothing to Predict Cloud Resource Provisioning

Ashraf A. Shahin[1,2]

[1]College of Computer and Information Sciences,
Al Imam Mohammad Ibn Saud Islamic University (IMSIU)
Riyadh, Kingdom of Saudi Arabia

[2]Department of Computer and Information Sciences, Institute of Statistical Studies & Research,
Cairo University,
Cairo, Egypt

*Abstract*—Elasticity is one of the key features of cloud computing that attracts many SaaS providers to minimize their services' cost. Cost is minimized by automatically provision and release computational resources depend on actual computational needs. However, delay of starting up new virtual resources can cause Service Level Agreement violation. Consequently, predicting cloud resources provisioning gains a lot of attention to scale computational resources in advance. However, most of current approaches do not consider multi-seasonality in cloud workloads. This paper proposes cloud resource provisioning prediction algorithm based on Holt-Winters exponential smoothing method. The proposed algorithm extends Holt-Winters exponential smoothing method to model cloud workload with multi-seasonal cycles. Prediction accuracy of the proposed algorithm has been improved by employing Artificial Bee Colony algorithm to optimize its parameters. Performance of the proposed algorithm has been evaluated and compared with double and triple exponential smoothing methods. Our results have shown that the proposed algorithm outperforms other methods.

## I. INTRODUCTION

Elasticity feature plays an important role in cloud computing by allowing SaaS providers to allocate and deallocate resources to their running services according to the demand. Elasticity allows SaaS providers to pay only for resources that are used by their cloud services [1]. However, the delay between requesting new resources and it being ready for use violates Service Level Agreement [2]. Therefore, forecasting future resource provisioning is needed to request resources in advance.

Exponential Smoothing is a very popular smoothing method and has been used through years in many forecasting situations [3]. Many researchers have exploited Exponential smoothing methods to predict future resource provisioning for cloud computing applications [4][5]. However, most of them have used double exponential smoothing, which cannot model workloads if there are seasonalities.

Most of cloud-computing applications' workloads are influenced by seasonal factors (e.g., day, week, month, year) and have more than one seasonal pattern [6][7][8]. Workload has intraday seasonal pattern if there is a similarity of request when comparing requests of the corresponding hour from one day to the next day. Intraweek seasonal pattern exists if there is a similarity between requests in two corresponding days from two adjacent weeks [3]. Therefore, there is a strong demand to use predictive approach that is able to capture all seasonality patterns.

This paper proposes resource usage prediction algorithm, which extends Holt-Winters exponential smoothing (HW) method to model multiple seasonal cycles. However, modeling multiple seasonal cycles requires large number of observation values. For example, predicting resource usage with intraday, intra-month, and intra-year seasonality patterns requires at least two years observation values. Moreover, finding optimal parameter values (smoothing constant, trend-smoothing constant and seasonal-smoothing constants) for multiple seasonality model is not an easy task.

Therefore, the proposed algorithm detects seasonality patterns from available historical data by applying seasonality test, and extends HW accordingly to model detected seasonality patterns. While historical data size grows up and more seasonality patterns are detected, HW is gradually extended to be able to model detected seasonality patterns. Furthermore, prediction accuracy of the proposed algorithm has been enhanced by using artificial bee colony algorithm to find near optimal values for its parameters. Thus, unlike most of current resource prediction approaches, the proposed algorithm does not require any minimum number of observations values before applying it. However, good prediction accuracy will not be achieved until several steps have been made.

The proposed algorithm has been evaluated using CloudSim simulator with real Web server log called Saskatchewan Log [6]. Performance of the proposed algorithm has been compared with double and triple exponential smoothing methods. Experimental results have shown that the proposed algorithm outperforms algorithms that use double or triple exponential smoothing methods.

This paper is organized as follows. In Section II related works are overviewed. The proposed algorithm is presented in Section III. Performance of the proposed algorithm is evaluated in Section IV. Finally, Section V concludes.

## II. RELATED WORK

The problem of predicting resource provisioning in cloud computing has been studied extensively over the last few years. Several prediction techniques have been used to predict cloud resource provisioning. However, most of current approaches do not consider multi-seasonality in cloud workload, and most of them use prediction techniques that do not have ability to model more than one seasonal cycle [4][9][10][11].

Islam et al. [12] have proposed framework to predict future resource usage in the cloud. The proposed framework uses two machine-learning algorithms (Neural Network and Linear Regression) with sliding window and cross validation techniques to predict cloud resource usage. The proposed framework is evaluated by using dataset that is collected by using TPC-W benchmark. Statistical metrics is proposed to assess prediction accuracy. However, the proposed framework uses three layers feed-forward Neural Network, which does not able to predict resource utilization when there are long time lags between events. Moreover, the proposed framework is tested with data that are collected from 135 minutes, which does not contain any seasonality. Therefore, prediction with seasonality is not examined.

Kanagala and Sekaran [4] have proposed dynamic threshold-based auto-scaling approach that considers virtual resource start-up and stabilization delays. Virtual resource utilization is predicted by using double exponential smoothing method, thresholds are adapted based on the predicted resource utilization to minimize violation of Service Level Agreement. However, double exponential smoothing method cannot be used to model seasonality.

In [5], Huang et al. have proposed resource utilization prediction model based on double exponential smoothing method. Prediction accuracy of the proposed model has been evaluated using CloudSim simulator, which shows that double exponential smoothing has better prediction accuracy than simple mean based method and weighted moving average method. However, smoothing constant and trend-smoothing constant are determined using trial method, which does not grant quality of the final solution.

Although, seasonal linear regression can be used to predict workload with seasonality, most of current approaches do not consider cloud workload seasonalities and use conventional linear regression to predict cloud resource utilization [1][13][14][15]. In [16], Yang et al. have proposed cost-aware auto-scaling approach, which predicts workload using linear regression model. The problem has been formulated as integer programming problem and solved using greedy heuristic to reduce costs. The proposed approach uses vertical and horizontal scaling methods. Allocated resources are scaled vertically by creating virtual machines on the same cluster node or using unallocated resources available at a particular cluster node to scale up a VM executing on it. Horizontal scaling is used to create virtual machines on other cluster nodes.

To gain benefits from several time series prediction models, Messias et al. [2] have proposed cloud workload prediction methodology that combines several time series forecasting models using genetic algorithm. Each time series prediction model has been assigned a weight, and genetic algorithm adapts the assigned weights to find the best weight combination that maximizes prediction accuracy.

Wei and Blake [17] have proposed an algorithm to predict future resource requirement in the cloud. The proposed algorithm uses five prediction models and differentiates between these models using root-mean-square-error (RMSE). Prediction model with the lowest RMSE is used to predict future resource requirement. Although, the proposed algorithm uses prediction techniques that do not have ability to model seasonality, it can be extended to include more prediction techniques with the ability to model seasonality.

Salah et al. [18] have proposed analytical model based on Markov chains to predict minimal number of VMs and load balancers required to satisfy Service Level Agreement such as throughput and response time. The proposed model has been validated using experimental testsbed deployed on the Amazon Web Services. Discrete-event simulation has been used to verify correctness of the proposed model.

## III. PROPOSED ALGORITHM

Although many researchers have employed double exponential smoothing for forecasting cloud applications' workload [5][4], double exponential smoothing does not able to model seasonality [3]. HW can be used for forecasting seasonal workloads [3]. However, HW is only able to model workloads with one seasonal pattern and cloud applications' workloads may have more than one seasonal pattern (e.g., intraday, intraweek, intra-month, intra-quarter, intra-year).

Therefore, in this paper, HW has been extended to be able to accommodate multi-seasonal patterns. As shown in equations 1-5, HW has been extended by adding seasonal indices and smoothing equation for each seasonal pattern.

$$S_t = \alpha \frac{X_t}{M(0)} + (1-\alpha)(S_{t-1} + B_{t-1}), \ 0 < \alpha \leq 1 \quad (1)$$

$$B_t = \beta(S_t - S_{t-1}) + (1-\beta)B_{t-1}, \ 0 < \beta \leq 1 \quad (2)$$

$$I_{i,t} = \gamma_i \frac{X_t \, I_{i,t-L_i}}{S_t \, M(0)} + (1-\gamma_i)I_{i,t-L_i}, \ 0 < \gamma_i \leq 1 \quad (3)$$

$$M(k) = \begin{cases} \prod_{i=1}^{n} I_{i,t-L_i+k}, & if \ n \geq 1 \\ 1, & if \ n = 0 \end{cases} \quad (4)$$

$$\hat{X}_t(k) = (S_t + k \, B_t) \, M(k) \quad (5)$$

where $t$ is an index denoting a time period, $S_t$ is smoothed value at time $t$, $X_t$ is observed value at time $t$, $B_t$ is trend factor at time $t$, $I_{i,t}$ is seasonal indices for seasonality pattern $i$, $L_i$ is number of periods in a completed seasonal cycle for seasonality pattern $i$, $\alpha$ is the smoothing constant, $\beta$ is the trend-smoothing constant, $\gamma_i$ is seasonal-smoothing constant for seasonality pattern i, n is number of seasonality patterns, and $\hat{X}_t(k)$ is the k-step-ahead forecast at time t.

Initial smoothed value for the level, $S_1$, is calculated as average of the first $2L_1$ periods, which are periods in the first two cycles from the first seasonal pattern. If there is no seasonality patterns, $S_1$ is initialized by the first observation value $X_1$. Initial value for the trend factor, $T_1$, is calculated as $1/L_1$ of average of the difference between first $L_1$ observations and second $L_1$ observations. If there is no seasonality patterns, $T_1$ is initialized as a difference between second observation value and first observation value ($X_2 - X_1$).

For each seasonal cycle, at least three completed seasonal data are required to initialize its seasonal indices. Seasonal indices are initialized as average of ratios of observed value to its centered moving average (calculated from $L_i$ periods around observed value), taken from the corresponding period in each of the first two completed seasonal data, which starts from $t = L_i/2$. For example, seasonal indices for seasonality pattern $i$ are calculated as following:

$$I_{i,t} = \left(\frac{X_t}{A_t} + \frac{X_{t+s_1}}{A_{t+s_1}}\right)/2 \;, t = \frac{L_i}{2}, \frac{L_i}{2}+1, \frac{L_i}{2}+2, \ldots, \frac{L_i}{2}+L_i$$

where $A_j$ is centered moving average around $X_j$ for $L_i$ periods

$$A_j = \sum_{i=j-L_i/2}^{j-1+L_i/2} X_i \Big/ L_i$$

Finally, artificial Bee colony algorithm is applied to determine near optimal values for smoothing constant, trend-smoothing constant and seasonal-smoothing constants that minimize Mean Squared Error (MSE).

$$\text{MSE}_t = \frac{1}{t}\sum_{i=1}^{t}(\hat{X}_i(k) - X_{i+k})^2$$

Algorithm 1 shows steps of the proposed algorithm. The first input is initial list of observation values that contains 60 observation values (from 60 minutes). This number of observation values is specified to start with enhanced accuracy. The second input is the list of completed seasonal cycles' length of expected seasonal patterns. Instead of applying seasonality test periodically, the second input specifies time points to test existence of seasonality patterns. The outputs are list of predicted values and list of seasonal cycles' length of detected seasonal patterns.

In the first line, initial smoothed value $s_1$ is set to the observed value $x_1$, and initial trend factor $b_1$ is set to ($x_2 - x_1$). Best values for smoothing constant α and trend-smoothing constant β are obtained by using Bee Colony Algorithm (Algorithm 2). At this point, number of seasonal cycles $n = 0$. Therefore, equations 1-5 are minimized to the following equations, which represent equations associated with Double Exponential Smoothing.

$$S_t = \alpha\, X_t + (1-\alpha)(S_{t-1} + B_{t-1}), \;\; 0 < \alpha \leq 1 \quad (6)$$

$$B_t = \beta(S_t - S_{t-1}) + (1-\beta)B_{t-1}, \;\; 0 < \beta \leq 1 \quad (7)$$

$$\hat{X}_t(k) = S_t + k B_t \quad (8)$$

Therefore, prediction accuracy of the proposed algorithm during the interval from $t = 61$ to $t = 3 * L_1 - 1$ (where $L_1$ is the number of periods in completed seasonal cycle for the first seasonal pattern) is very similar to prediction accuracy of double exponential smoothing.

If $t$ equals to $3 * l'_i$, where $l'_i \in L'$, seasonality test is applied to check if the list of observed values $X$ has seasonal pattern with length $l'_i$ or not. New seasonal pattern is detected if autocorrelation coefficient is greater than or equal 0.3. Length of the detected seasonal pattern $l'_i$ is added to the list of seasonal cycles' length $L$, and number of detected seasonal patterns $n$ is increased. List of seasonal indices for the new seasonal pattern is calculated and added to $I$. Smoothing constant, trend-smoothing constant, and seasonal-smoothing constants are updated to the near optimal values using artificial Bee colony algorithm. Finally, extended formula is

---

ALGORITHM 1: The proposed algorithm

**INPUTS**:
 $X$: initial list of observation values
 $L'$: list of expected seasonal cycles' length

**OUTPUTS**:
 $\hat{X}(k)$: list of $k$-step-ahead predicted values
 $L$: list of seasonal cycles' length

**Begin**
1: $s_1 = x_1$
2: Initialize $S$ and add $s_1$ to $S$
3: $b_1 = x_2 - x_1$
4: Initialize trend factor list B and add $b_1$ to B
5: Get Best Constants Using Bee Colony Algorithm
6: $n = 0$, where n is the number of seasonal cycles in $X$
7: $t = 1$, where t is an index denoting a time period
8: **while** $t \leq 60$
9:    Calculate $s_t$ using equation 1 and add it to $S$
10:    Calculate $b_t$ using equation 2 and add it to $B$
11:    Calculate $\hat{X}_t(k)$ using equation 5 and add it to $\hat{X}(k)$
12:    $t++$
13: **end while**
14: **for** each new observation value $x_t$ at time t
15:    Add $x_t$ to X
16:    **if** $t/3 \in L'$
17:       Apply seasonality test
18:       **if** autocorrelation coefficient$\geq 0.3$
19:          $n++$
20:          Add $t/3$ to $L$
21:          Initialize seasonal indices list $I_n$ for seasonal cycle with length $t/3$
22:          Get Best Constants Using Bee Colony Algorithm
23:       **end if**
24:    **end if**
25:    Calculate $s_t$ using equation 1 and add it to $S$
26:    Calculate $b_t$ using equation 2 and add it to $B$
27:    Calculate $I_{i,t}$ using equation 3 for all $i = 1,2,\ldots,n$ and add it to $I$
28:    Calculate $\hat{X}_t(k)$ using equation 5 and add it to $\hat{X}(k)$
29: **end for**
30: **return**
**End**

---

employed to predict future required resources.

Algorithm 2 shows steps of determining best values for smoothing constant α, trend-smoothing constant $\beta$, and seasonal-smoothing constants $\gamma$ by using artificial Bee colony optimization algorithm.

At the beginning, initial population $P$ is initialized with $ns$ scout bees, which are randomly scattered across solution space. Here, all constants (smoothing constant α, trend-smoothing constant $\beta$, and seasonal-smoothing constants $\gamma$) are greater than zero and less than or equal one. For each scout bee in $ns$, flower patch is delimited that contains its neighborhood.

MSE is calculated for each scout bee by applying equations 1-5 from $t = l_n$ to $t = \|X\|$, where $l_n$ is number of periods in completed seasonal cycle for the largest seasonal pattern. if $n = 0$, MSE is calculated by applying equations 1-5 from $t = 2$ to $t = \|X\|$. Scouts are sorted in ascending order according to their MSE.

Best sites $nb$ with lowest MSE are selected from $ns$, and elite sites $ne$ with most lowest MSE are selected from $nb$.

Each scout in $nb$ performs waggle dance to recruit forager bees to search further in its flower patch. Such that, number of

---
ALGORITHM 2: Determine Best Constants Using Bee Colony Algorithm
---

**INPUTS**:
  $S$: list of smoothed values
  $X$: list of observed values
  $B$: trend factor list
  $I$: seasonal indices list
  $n$: number of detected seasonality patterns
  $L$: list of seasonal cycles' length
  $MaxIter$: maximum iteration number
  $MaxError$: maximum allowed error

**OUTPUTS**:
  Near optimal values for smoothing constant α, trend-smoothing constant $\beta$, and seasonal-smoothing constants $\gamma$

**Begin**
  1: Generate initial population $P$ with ns scout bees
  2: Specify flower patch for each scout in $nb$
  3: Calculate MSE for each scout in $P$
  4: Sort scouts in $P$ in ascending order based on their MSE values
  5: $i = 0$
  6: **while** $i \leq MaxIter$ or
         $FitnessValue_i - FitnessValue_{i-1} \leq MaxError$
  7:    $i++$
  8:    Select best sites $nb$ from $ns$
  9:    Select elite sites $ne$ from $nb$
  10:   Recruit forager bees to $ne$ and $nb - ne$
  11:   Apply local search to find fittest bee of each flower patch
  12:   Generate random solutions for non-best sites $ns - nb$
  13:   Calculate *MSE* for non-best sites $ns - nb$
  14:   Sort all scouts in ns in ascending order based on their MSE values
  15:   $FitnessValue_i = MSE$ of the first scout in the sorted $ns$
  16: **end while**
  17: Determine constants' value according to fittest scout in population $P$
  18: **return**
**End**

---

recruited forager bees to $ne$ ($nre$) is greater than number of recruited forager bees to the remaining best sites $nb - ne$ ($nrb$).

To find fittest bee of each flower patch, recruited forager bees are randomly distributed in flower patch. MSE is calculated for each bee. If there is recruited forager bee with MSE lower than MSE of its scout bee, fittest bee will be selected as a new scout. Otherwise, flower patch will be shrunken around its scout. After pre-specified number of search cycles, the fittest bee of each flower patch is returned as a local optimal solution.

New solutions are generated randomly for non-best sites $ns - nb$, and all scout in $ns$ are sorted in ascending order according to their MSE. This search cycle will be repeated until reaching termination condition. Finally, values of smoothing constant α, trend-smoothing constant $\beta$, and seasonal-smoothing constants $\gamma$ are obtained from fittest scout bee in current population.

## IV. PERFORMANCE EVALUATION

To evaluate performance of the proposed algorithm, its performance have been compared with double and triple exponential smoothing methods. The following subsections, describe evaluation environment settings and discuss simulations' results.

### A. Evaluation environment settings

The proposed algorithm has been evaluated using real Web server log called Saskatchewan Log [6]. Saskatchewan log contains HTTP requests to the University of Saskatchewan's WWW server, which is located in Saskatoon, Saskatchewan, Canada. This log was collected from 00:00:00 June 1, 1995 to 23:59:59 December 31, 1995, a total of 214 days [6].

Cloudlets have been generated according to Saskatchewan log and sent to CloudSim simulator. For each minute, CloudSim simulator calculates total required CPU to process incoming requests without violating Service Level Agreement. The proposed algorithm receives required CPU as observed value and predicts required CPU after $k$-minutes. $K$ has been set to 15, where $k$ is a virtual machine startup delay.

To evaluate accuracy of the proposed algorithm, three evaluation metrics have been used:

- *Mean absolute percentage error (MAPE),* which is defined as following:

$$MAPE_t = \frac{1}{t} \sum_{i=1}^{t} \frac{|\hat{X}_t(k) - X_{t+k}|}{X_{t+k}}$$

where $MAPE_t$ is mean absolute percentage error at time t, $\hat{X}_t(k)$ is the $k$-step-ahead forecast at time $t$, and $X_{t+k}$ is observed value at time $t + k$. A smaller value of $\text{MAPE}_t$ implies a better prediction accuracy.

-*Percentage of predictions within 25% ($PRED(25)$),* percentage of prediction within 25% at time $t$ is defined as following:

$$PRED(25)_t = \frac{1}{t} \left\| \left\{ \hat{X}_i(k) : \frac{|\hat{X}_i(k) - X_{i+k}|}{X_{i+k}} < 25\%, \quad 0 \leq i \leq t \right\} \right\|$$

$PRED(25)_t$ values are between 0 and 1. Prediction will be more effective if $PRED(25)_t$ value is closer to 1.

-*Root Mean Squared Error* (RMSE), RMSE at time $t$ is defined as following:

$$\mathrm{RMSE}_t = \sqrt{\frac{1}{t}\sum_{i=1}^{t}(\hat{X}_i(k) - X_{i+k})^2}$$

A smaller value of $\mathrm{RMSE}_t$ implies better prediction accuracy.

### B. Evaluation results

Although, the proposed algorithm has been evaluated with many real workload traces such as [6][7][8] in this evaluation only one of them has been shown, which is Saskatchewan-http.

Fig. 1 compares the proposed multi-seasonal algorithm with double and triple exponential smoothing methods using Mean Absolute Percentage Error (MAPE), which has been defined in the previous section. As shown in Fig. 1, MAPE of the proposed multi-seasonal algorithm stays below 29% while triple and double are above 44% and 135% respectively. Fig. 2 shows that more than 57% of predicted values by using the proposed multi-seasonal algorithm are with prediction error less than 25%. In another side, 38% of triple exponential smoothing predictions are within 25%, and 8-18% of double exponential smoothing predictions are within 25%. Finally, Root Main Square Error of the proposed multi-seasonal algorithm has been compared with double and triple exponential smoothing methods in Fig. 3, which shows that RMSE of the proposed multi-seasonal algorithm is better than other methods.

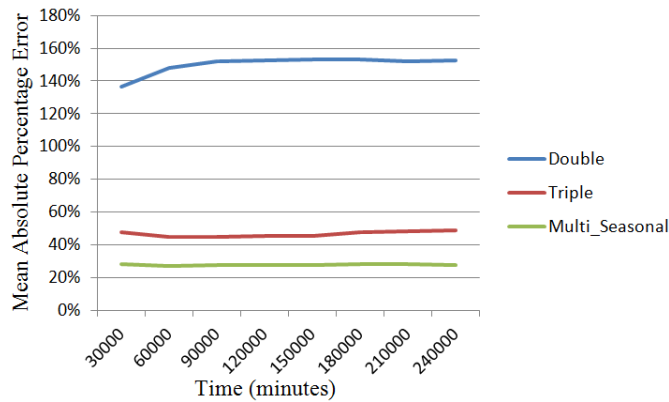

Fig. 1.  Mean Absolute Percentage Error comparison

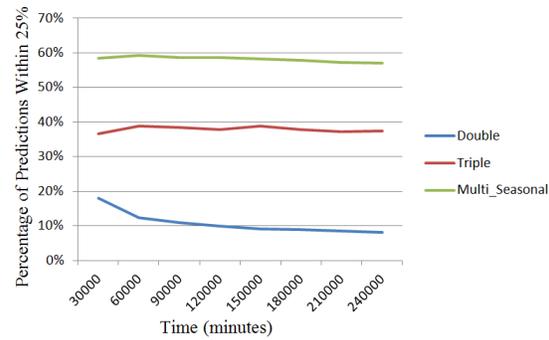

Fig. 2.  Percentage of Predictions Within 25% comparison

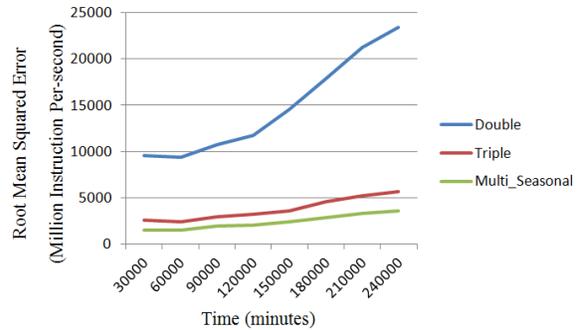

Fig. 3.  Root Main Square Error comparison

## V. CONCLUSION

This paper has proposed predictive algorithm to predict cloud resource provisioning. According to available historical data and detected seasonal cycles, Holt-Winters exponential smoothing method has been extended to allow modeling multiple seasonal cycles with minimum number of observation values. Artificial Bee Colony algorithm has been exploited to find near optimal parameters value for the proposed algorithm. Prediction accuracy of the proposed algorithm has been evaluated by using CloudSim simulator with real workload called Saskatchewan-http. Our results have shown the effectiveness of the proposed algorithm among other methods. Finally, the paper concludes that modeling multiple seasonal cycles during predicting cloud resource provisioning is an essential step toward accurate cloud resource prediction.

As future work, long short-term memory recurrent neural networks will be incorporated with the proposed algorithm to predict cloud resource utilization when there are very long and variant time lags between events. Because, in seasonality patterns, seasonal cycle length is considered constant for each seasonal pattern. However, in some cases, lags between events are variant and have to be considered during prediction.